\documentclass{ws-mpla}

\usepackage[super]{cite}
\usepackage{xcolor}
\usepackage[verbose,hypertexnames=false]{hyperref}
\definecolor{blue}{rgb}{0.0, 0.0, 1.0}
\definecolor{red}{rgb}{1.0, 0.0, 0.0}
\definecolor{royalblue}{rgb}{0.0, 0.14, 0.4}
\hypersetup{colorlinks=true,linkcolor=red, citecolor=blue, urlcolor=blue}
\def\orcid#1{\kern .08em\href{https://orcid.org/#1}{\includegraphics[keepaspectratio,width=0.7em]{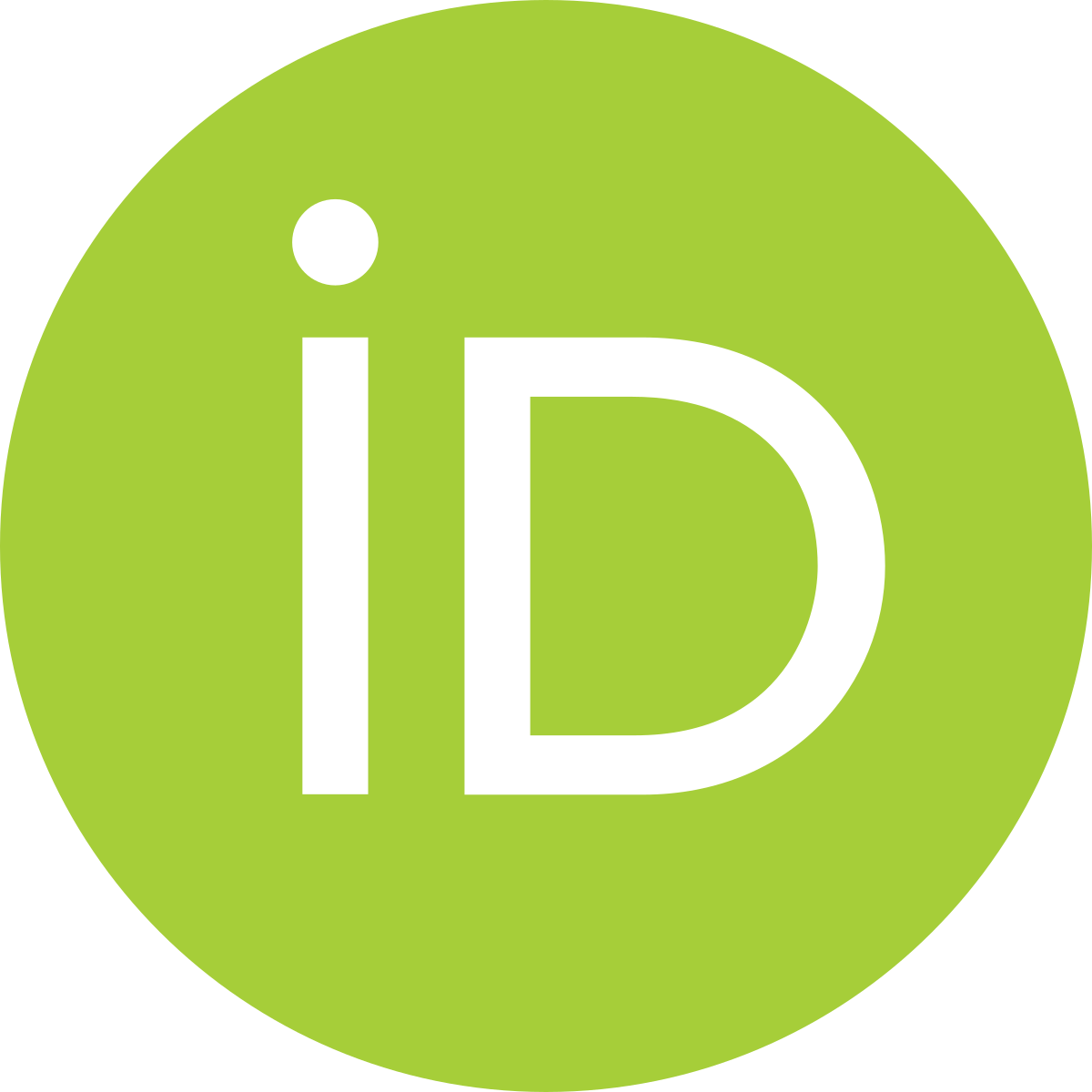}}}
\usepackage[normalem]{ulem}
\usepackage{booktabs}

\usepackage[normalem]{ulem}
\usepackage{xcolor}

\begin{document}

\markboth{Baihaqi \textit{et al.,}}{Determination of HERA coherent diffractive $J/\Psi$ production cross section via 
ANN}

\catchline{}{}{}{}{}

\title{Determination of the HERA coherent diffractive $J/\psi$ production cross section via artificial neural network}

\author{Taufiq I.~Baihaqi\orcid{0009-0008-3148-6994}}
\address{Department of Physics, Universitas Gadjah Mada,  BLS 21, Indonesia\\
Research Center for Quantum Physics, National Research and Innovation Agency (BRIN),\\
South Tangerang 15314, Indonesia\\
E-Mail Address: taufiqiqbalbaihaqi@mail.ugm.ac.id}

\author{Chalis~Setyadi\orcid{0000-0002-5853-4238}}
\address{Department of Physics, Universitas Gadjah Mada, BLS 21, Indonesia\\
E-Mail Address: chalis@mail.ugm.ac.id}

\author{Zulkaida~Akbar\orcid{0000-0002-5373-6121}}
\address{Research Center for Quantum Physics, National Research and Innovation Agency (BRIN),\\
South Tangerang 15314, Indonesia\\
E-Mail Address: akbar.zulkaida@gmail.com}

\author{Parada~T.~P.~Hutauruk \orcid{0000-0002-4225-7109}}
\address{Department of Physics, Pukyong National University (PKNU) Busan 48513, Korea \\
Departemen Fisika, FMIPA, Universitas Indonesia
Depok, 16424, Indonesia\\
E-Mail Address: phutauruk@gmail.com}

\author{Apriadi~Salim~Adam\orcid{0000-0001-6587-5156}}
\address{Research Center for Quantum Physics, National Research and Innovation Agency (BRIN),\\
South Tangerang 15314, Indonesia\\
E-Mail Address: apriadi.salim.adam@brin.go.id}


\maketitle


\begin{abstract}
An exclusive coherent diffractive $J/\psi$ production dataset from HERA, covering a large kinematic range in the photon virtuality $Q^2$, the squared momentum transfer $t$, and the photon-proton center-of-mass energy $W$, has been analyzed using various theoretical models with different approaches. In common model analyses, the inherent assumptions and limited kinematic applicability somewhat restrict the predictive power of the models, resulting in model-dependent prediction results. In this paper, we present our model-independent approach for the same reaction process and dataset, utilizing an artificial neural network (ANN) technique. The prediction of the best ANN model for the HERA differential cross-section dataset over a range of $W$, $Q^2$, and $t$ is obtained. We then extend the ANN model by combining the HERA and LHC data at various values of $W$ to predict the total photoproduction cross-section and demonstrate how to extract the exponential slope $b$. We find that the exponential slope $b$ strongly depends on $Q^2$ and $W$.
\end{abstract}

\section{Introduction 
}
Exclusive~\cite{Lepage:1980fj} coherent diffractive $J/\psi$ production serves as an excellent probe for high-energy physics phenomenology due to its sensitivity to the target profile, which enables us to gain new insights into the fundamental properties of the hadron. Such a $J/\psi$ production process has recently gathered significant attention since it is one of the Electron-Ion Collider (EIC) physics programs and will be planned to measure in future modern experiments with a high precision like the EIC~\cite{Accardi:2012qut,Li:2022kwn}, the EIC in China (EicC)~\cite{Anderle:2021wcy}, and the Large Hadron Electron Collider (LHeC)~\cite{LHeCStudyGroup:2012zhm}. For details and a comprehensive literature review about this process, we refer the interested readers to see Refs.~\cite{Ryskin:1992ui,Brodsky:1994kf,Frankfurt:1995jw,Collins:1996fb,Boer:2024ylx}. On the other hand, extensive data on $J/\psi$ production, spanning a wide range of photon virtualities ($Q^2 = -q^2$) and squared momentum transfers ($t$), have been collected from HERA experiments~\cite{H1:2000kis,H1:2005dtp, H1:2013okq,ZEUS:2002wfj}.
These collected datasets enable us to evaluate various theoretical models using different approaches and to pinpoint key phenomena, such as the saturation effect~\cite{Golec-Biernat:1999qor,Golec-Biernat:1998zce,Gelis:2010nm,Kowalski:2003hm,Weigert:2005us}, the proton structure~\cite{Amoroso:2022eow}, and the geometric shape of the proton~\cite{Schlichting:2014ipa,Mantysaari:2016ykx,Mantysaari:2016jaz}. In addition, the results of the elastic photoproduction of light vector mesons in lepton--proton collisions at HERA have verified the expected universal Regge behavior~\cite{H1:1996prv,ZEUS:1995bfs}, which predicts a power-law growth of the cross section $\sigma \propto W^\delta$, where $\delta$ determines the energy dependence~\cite{Donnachie:1994zb}. In the case of heavy quarkonia, such as $J/\psi$, the cross section exhibits a significantly steeper rise~\cite{H1:2005dtp}. Unlike the light vector meson case, this rapid increase of the power-law behavior has been confirmed by perturbative QCD (pQCD) models, demonstrating a strong sensitivity to the gluon distribution in the proton. It is worth noting that the slope parameter $\delta$ does not change significantly with the $Q^2$.

In the literature, the existing theoretical models that describe the HERA~\cite{H1:2000kis,H1:2005dtp, H1:2013okq,ZEUS:2002wfj,ZEUS:2004yeh} and LHC~\cite{Bursche:2018eni,LHCb:2018rcm} data on the exclusive coherent diffractive processes mostly rely on the dipole picture approach, where, in this approach, the photon splits up into a quark-antiquark ($q\bar{q}$) pair and then it interacts with the proton target~\cite{Iancu:2017fzn,Kowalski:2003hm,Golec-Biernat:1998zce,Mueller:1994jq}. Despite the success of this approach in a widely applied framework, several significant challenges remain, which are related to the uncertainties in various model component variables, namely the vector meson wave function, the target profile, the skewness correction, and the real-imaginary part of the scattering amplitude correction~\cite{Kowalski:2006hc}. In addition, most theoretical models are constrained by narrow kinematic ranges, typically at small $t$ and moderate $Q^2$, making it difficult to describe the full spectrum of the data. This becomes an interesting avenue for study and deserves further investigation.

In improving the accuracy of the model predictions, there has been increasing interest in utilizing machine learning for analyzing large datasets in high-energy physics research. Among them, the most commonly used methods are Boosted Decision Trees (BDTs) and Artificial Neural Networks (ANNs)~\cite{Albertsson:2018maf}. Compared to tree-based methods such as boosted decision trees, ANNs provide smooth and differentiable functional representations, which are essential for extracting derived observables such as the exponential $t$-slope parameter $b$.

In this work, we explore whether a data-driven approach using the ANN can help us tackle such a theoretical model limitation, particularly in reducing model dependencies on the specific theoretical model assumptions and in addressing the multi-dimensional correlations in data. To achieve this, we train three distinct ANN with multiple epoch and iterative learning using H1 and ZEUS data from HERA experiment for the differential cross-section ($d\sigma/dt$) and we assess their capability to predict the differential and total ($\sigma$) cross-sections across the ranges of $Q^2$ and $W$ values available from the HERA and the LHC, focusing on the exponential dependence of $d\sigma/dt$. We expect that this approach can minimize the reliance on the uncertain model variables and provide a data-driven ANN framework for evaluating the cross-section. 

This paper is organized as follows. Section~\ref{sec:II} briefly describes the theoretical framework for the differential cross section of exclusive coherent diffractive $J/\psi$ production, along with the main ingredients on which the cross section depends and which are commonly modeled. In Sec.~\ref{sec:III}, we present the ANN algorithms used in the present work, including training, validating, and testing the ANN models. We also check the model losses and the accuracy of the ANN models, as well as opting for the best and most reliable ANN model for predicting the differential and total cross-section from the HERA, LHC, and EIC experimental data. In Sec.~\ref{sec:IV}, we present the prediction results from the best and most reliable ANN model. Finally, section~\ref{sec:V} is devoted to a summary and outlook.

\section{Exclusive coherent diffractive $J/\psi$ production}\label{sec:II}
In this section, we provide a brief description of the theoretical framework for the exclusive coherent diffractive reaction process. The exclusive coherent diffractive production of the $J/\psi$ meson in photon–proton ($\gamma p$) scattering involves the exchange of a color-singlet gluonic system, commonly referred to as the “Pomeron,” which leaves the proton intact. The final state is characterized by a $J/\psi$ meson and a proton separated by a large rapidity gap. The process is described by the reaction process of $\gamma^{(*)} p \to J/\psi\, p$. 
The central observable in this reaction process is the differential cross section $d\sigma/dt$. In the small $t$, the differential cross section typically exhibits an exponential form, and it can be defined as
\begin{eqnarray}
\label{eq1}
    \frac{d\sigma}{dt} \propto \exp\big[-b|t|\big],
\end{eqnarray}
where the slope parameter $b$ encodes information about the transverse spatial distribution of gluons of the proton (nuclei) target in the transverse plane. If the impact parameter enters the cross section via a purely Gaussian proton shape function $T(\mathbf{b}_\perp)$, the slope parameter $b$ is directly determined by the width parameter $B_G$, where $T(\mathbf{b}_\perp) \propto \exp(-\mathbf{b}_\perp^2/B_G)$. This function describes the transverse gluonic charge distribution inside the proton. This relationship arises because the square of the momentum transfer $t$ is the conjugate variable to the impact parameter $\mathbf{b}_\perp$, related through a two-dimensional Fourier transform~\cite{Kowalski:2006hc}.

A general formula for the differential cross section for vector meson $V$ production in the reaction process $\gamma^{(*)}p \rightarrow Vp$ can be expressed as~\cite{Kowalski:2006hc,Boer:2023mip}
\begin{eqnarray}
    \frac{d\sigma^{\gamma^{(*)}p\rightarrow Vp}}{dt} &=& \frac{1}{16\pi} \left|{\mathcal{A}^{\gamma^{(*)}p\rightarrow Vp}}\right|^2,
    \label{dsdt}
\end{eqnarray}
where the scattering amplitude $\mathcal{A}^{\gamma^{(*)}p\rightarrow Vp}$ is defined as
\begin{eqnarray}
    \mathcal{A}^{\gamma^{(*)}p\rightarrow Vp} &=& 2i \int d^2\bm{r}_{\perp} \int_0^1 \frac{dz}{4\pi} (\Psi_V^* \Psi_\gamma)\nonumber \\
    &\times& \int d^2\bm{b}_{\perp} \exp \big[-i\big(  \bm{b}_{\perp} - \big(\frac{1}{2}-z\big)\bm{r}_{\perp}\big)\cdot\Delta_{\perp}\big] \hat{\sigma}(x,\bm{r}_{\perp},\bm{b}_{\perp}).
    \label{elasticscatteringamplitude}
\end{eqnarray}

From Eq.~(\ref{elasticscatteringamplitude}), it is clearly shown that the scattering amplitude is influenced by the overlap between the photon wave function $\Psi_\gamma$ and the vector meson wave function $\Psi_V$, and the dipole–proton cross section $\hat{\sigma}(x,\bm{r}_{\perp},\bm{b}_{\perp})$. The photon wave function can be calculated perturbatively, while the vector meson wave function requires phenomenological modeling. Several models are available for this purpose, such as those proposed in Refs~\cite{Dosch:1996ss,Nemchik:1996cw,Kowalski:2006hc,Kowalski:2003hm,Kopeliovich:1991pu}. The dipole–proton cross section encodes dependencies on the target profile and the saturation effect, which may depend on the impact parameter $\bm{b}_{\perp}$, the dipole size $\bm{r}_{\perp}$ and orientation, and the parton longitudinal momentum fraction $x$ that carries from the proton momentum. 

Among other model approaches, the McLerran--Venugopalan (MV) model~\cite{McLerran:1993ni,McLerran:1993ka} is a widely used approach that describes the color charge distribution of the target to compute the dipole--proton cross section. In the small-$x$ regime, the Golec-Biernat-W\"{u}sthoff (GBW)~\cite{Golec-Biernat:1998zce,Golec-Biernat:1999qor} and color glass condensate (CGC)~\cite{Iancu:2003ge} models are frequently employed to capture saturation effects; the latter was subsequently modified to include the impact parameter, known as the b-CGC model~\cite{Kowalski:2006hc}. Other dipole cross-section models are also available in the literature, such as those used in Ref.~\cite{Forshaw:2004vv,Kowalski:2003hm}.
Since the scattering amplitude is calculated from the purely imaginary part (optical theorem), a phenomenological correction is usually applied to account for the real part~\cite{Nemchik:1996cw,Forshaw:2003ki,Martin:1999wb}. Another correction associated with the skewedness effect is also commonly included in the cross-section \cite{Martin:1999wb,Shuvaev:1999ce}. The combination of both phenomenological corrections will vary the effects depending on the applied model, ranging from 10\% to 50\%. In addition, in some approaches, the QCD running coupling $\alpha_S(Q^2)$ is taken to depend on the dipole size~\cite{Lappi:2012vw}; otherwise, a fixed coupling constant is used for simplicity. 

All these dependencies enter through phenomenological modeling, with several parameters determined by fits to data over various observables and kinematic regions. Given the strong model dependence in such calculations, a data-driven framework employing an ANN serves as an effective tool to reduce such biases.
The ANN provides a data-driven framework capable of handling the complex feature space, capturing the nonlinear relationships between the input kinematic variables and observables, while also generalizing patterns across the cross-section data. The ANN framework is described in detail in Sec.~\ref{sec:III}.

\section{ANN model framework
\label{sec:III}}
In this section, we describe in detail the artificial neural network (ANN) framework employed to model the exclusive coherent diffractive $J/\psi$ production cross section. The objective of this framework is to construct a flexible, data-driven regression model capable of capturing the multidimensional and nonlinear dependence of the differential cross section on the kinematic variables ($Q^2$, $W$, $t$, $y$), while simultaneously providing a reliable and quantitative estimate of the associated uncertainties.
The ANN approach is particularly well suited for this task because it does not rely on explicit assumptions about the underlying hadronic dynamics (such as dipole amplitudes, proton profiles, or vector-meson wave functions), but instead learns the relevant correlations directly from experimental data. Moreover, compared to tree-based methods such as boosted decision trees, ANNs provide smooth and differentiable functional representations, which are essential for extracting derived observables such as the exponential $t$-slope parameter $b$.

\subsection{HERA Dataset and pre-processing 
\label{data preprocessing}}
The ANN model developed in the present work is trained using a combined dataset for the $J/\psi$ photo- and electroproduction process collected from the H1~\cite{H1:2005dtp} and ZEUS~\cite{ZEUS:2004yeh} collaborations measured at HERA experiments. The total dataset used in the analysis includes 108 data points, covering the kinematic values of $Q^2$ (photon virtuality) in the range $0.05$ GeV$^2<Q^2<$ 100 GeV$^2$,
photon-proton center-of-mass energy in the range 20 GeV $<W <$ 250 GeV, and $\mid t \mid <$ 1.2 GeV$^2$. To effectively capture the physical dynamics, four input variables used for each data point are $Q^2$, $W$, $t$, and $y$\footnote{In the lab frame, variable $y$ is the fraction electron energy loss of the incoming particle.}, with $y=(Q^2 + W^2)/s$ being the inelasticity and $\sqrt{s} = 318$ GeV is the HERA lepton-proton center of mass energy.

To improve numerical stability and learning efficiency, variables that span several orders of magnitude (Q², W, and y) are transformed to a logarithmic scale, resulting in the input feature set {log($Q^2$), log($W$), $t$, log($y$)}. The momentum transfer $t$ is kept in a linear scale because it can approach zero and already enters the cross section in an approximately exponential form, making a logarithmic transformation neither well defined nor necessary. The target variables (predictions) are the differential cross section $\frac{d\sigma}{dt}$, and its uncertainty, which are also transformed to logarithmic space. This transformation stabilizes the variance, reduces skewness in the target distribution, and improves convergence during training.

The experimental data errors ($\Delta\sigma$) are also processed and propagated to log-space. Note that the datasets differ in their error features, where H1 has symmetric statistical and systematic uncertainties, whereas ZEUS has symmetric statistical but asymmetric systematic uncertainties. To deal with such an error dataset features, we employ the Gaussian Negative Log Likelihood (NLL) loss function to have a single symmetry uncertainty input variable (See details in Sec.~\ref{loss function}). To do so, the experimental data errors can be processed through the following three steps:
\begin{enumerate}
    \item\textbf{Error combination}: For each data point, the statistical and systematic errors are added in quadrature (square root of the sum of squares). For the asymmetrical error of the ZEUS data, it will be performed separately for the upper and lower error limits.
    \item \textbf{Error symmetrization}: The asymmetric total data errors can be symmetrized by taking the arithmetic mean: $\Delta\sigma_\text{exp}=(\Delta\sigma_{\text{total up}}+\Delta\sigma_{\text{total down}})/2$, where $\Delta\sigma_{\text{total up}}$ and $\Delta\sigma_{\text{total down}}$ represent the quadrature sum of statistical and systematic errors for the upper and lower bounds, respectively. This approximation is adopted to maintain consistency with the Gaussian negative log-likelihood loss function, which assumes a symmetric conditional likelihood. Given that the asymmetries in the ZEUS data are moderate and that the ANN infers a smooth mean function constrained by multiple measurements, we expect the impact of this approximation on the predicted central values to be small.
    \item \textbf{Error propagation}: In this step, the symmetrical error $\Delta\sigma_\text{exp}$ is propagated to log-space to match the $\bar{y}_\text{true}$ target. This is done using standard error propagation: $\delta y_\text{err}=(\Delta\sigma_\text{exp}/\sigma)$. The result of $\delta y_\text{err}$ value is used as the uncertainty input $\delta_\text{exp}$ for the model's loss function.
    To ensure consistency with the logarithmic target variable $\bar{y}_\text{true}$, the symmetrical experimental uncertainty $\Delta\sigma_\text{exp}$ is propagated into log-space via first-order error propagation. This yields the fixed uncertainty term: $\delta_{\text{exp}} \approx \Delta\sigma_\text{exp}/\sigma_\text{exp}$. This term $\delta_{\text{exp}}$ is subsequently used as the fixed noise component in the loss function.
\end{enumerate}

Once all features, targets, and errors are processed, the dataset is randomly partitioned into three subsets: training, validation, and testing. To guarantee the reproducibility of the results, a fixed random seed is utilized throughout the splitting process. Specifically, 10\% of the total data is set aside as a test set to evaluate the model's generalization capability. The remaining data is then divided into 90\% for training and 10\% for validation. Before training, the input features are standardized using a \texttt{StandardScaler}. To prevent data leakage, the scaler is fitted exclusively on the training set and subsequently applied to transform both the validation and test sets.

\subsection{Model Architecture and Uncertainty Quantification}
\subsubsection{Heteroscedastic Architecture (Aleatoric Uncertainty)}

The architecture of the model is designed using the Keras functional API~\cite{Chollet:2015} to produce two outputs, such as the prediction and its uncertainty. The model accepts the 4 processed input features. The model's core consists of three hidden layers with 64, 64, and 32 neurons, respectively. The \texttt{tanh} activation function is used for each hidden layer. To prevent overfitting and improve generalization, L2 regularization with a penalty factor of $\lambda = 1\times10^{-4}$ is applied to each hidden layer~\cite{Krogh:1992}. By constraining the magnitude of the network weights, this regularization technique limits the model's complexity, thereby preventing it from fitting statistical noise and ensuring robust generalization. This architecture terminates in two separate linear output heads, designed for heteroscedastic regression~\cite{Nix1994EstimatingTM}:
\begin{itemize}
    \item \textbf{Prediction output ($\mu$)}: A single neuron predicting the mean of the target value of the logarithmic cross-section distribution, corresponding to 
    $\mu_\text{pred}=\log(\sigma)$.
    \item \textbf{Uncertainty output ($\log(\sigma^2_\text{model})$)}: A single neuron predicting the logarithm of the model's (aleatoric) variance.
\end{itemize}

Using these two outputs, it allows the model to not only predict the value of ($d\sigma/dt$) but also the confidence level of that prediction, which is known as aleatoric uncertainty.
The implementation of these two outputs enables the model to predict not only the central value of the cross-section but also the associated aleatoric uncertainty, which reflects the intrinsic noise of the data.

\subsubsection{Loss Function (Gaussian NLL)}\label{loss function}
 
To train the heteroscedastic model, the Gaussian NLL loss function is considered, following the framework of likelihood estimation for heteroscedastic regression~\cite{Yuan2004DoublyPL, Le2005Heteroscedastic}. This loss function measures how probable the true target value of $\bar{y}_{\text{true}}$ given the Gaussian distribution predicted by the model. The core of this method is the definition of the total variance ($\Delta\sigma^2_{\text{total}}$), which combines the fixed uncertainty from the experimental data and the learned aleatoric uncertainty from the model. The total variance is defined as follows.

\begin{equation}
    \Delta\sigma^2_{\text{total}} = \delta^2_{\text{exp}} + \Delta\sigma^2_{\text{model}},
\end{equation}
where $\delta^2_{\text{exp}}$ is the fixed experimental variance derived from error propagation (as defined in Sec.~\ref{data preprocessing}) and $\Delta\sigma^2_{\text{model}}$ is the variance predicted by the model's second output (aleatoric uncertainty).
The NLL loss function is then defined as

\begin{equation}
    \mathcal{L}_{\text{NLL}} = \frac{1}{2} \left( \log(2\pi\Delta\sigma^2_{\text{total}}) + \frac{(\bar{y}_{\text{true}} - \mu_{\text{pred}})^2}{\Delta\sigma^2_{\text{total}}} \right).
\end{equation}
By minimizing the loss function, the model simultaneously learns to predict the mean ($\mu_{\text{pred}}$) and the aleatoric uncertainty ($\Delta\sigma^2_{\text{model}}$) that yields the best fit to the data.

\subsubsection{Deep Ensemble (Epistemic Uncertainty)}\label{epistemic uncertainty}
A single neural network is susceptible to epistemic uncertainty, even when trained with a heteroscedastic loss function designed for aleatoric uncertainty~\cite{Kendall:2017tnb}. This uncertainty arises from the model itself, particularly its sensitivity to weight initialization. The necessity for a robust ensemble approach was established through preliminary stability tests using repeated $k$-fold cross-validation. In this procedure, the dataset is randomly partitioned into $k=5$ distinct folds. The process is repeated 10 times with different random seeds, resulting in a total of 50 independent training runs. In each run, the model is trained on $k-1$ folds and evaluated on the remaining fold. These comprehensive tests revealed that the model's performance fluctuates significantly with variations in data sampling, indicating high sensitivity to the training data composition, as illustrated by the $\chi^2/ndf$ and $\sigma_{\text{pull}}$ values in Fig.~\ref{fig:overfitting_avoidance}. We emphasize that this cross-validation procedure serves exclusively as a diagnostic stability test and is not employed for the final cross-section predictions. Instead, to rigorously quantify and isolate the uncertainty arising purely from model initialization, we implement the Deep Ensemble method~\cite{Lakshminarayanan:2016simple}.

Thus, we train $N = 100$ architecturally identical models. All models are trained exactly using the same (fixed) training and validation sets. The only variable changed for each training run is the random seed for weight initialization. This process involves 100 trained models that, while all are optimal, converge to slightly different local minima. The difference or disagreement in their predictions is a direct measure of epistemic uncertainty. After the $N=100$ models are trained, a single input $X$ will produce $N$ central value predictions ($\{\mu_1, \mu_2, \dots, \mu_N\}$) and $N$ aleatoric variance predictions ($\{\Delta\sigma^2_{\text{model},1}, \dots, \Delta\sigma^2_{\text{model},N}\}$). The final predicted value ($\mu_{\text{final}}$) is obtained by taking the mean of all central value predictions:
\begin{equation}
    \mu_{\text{final}}=\frac{1}{N}\sum_{i=1}^{N}\mu_i.
\end{equation}
The next step is the quantification of the final error bar, which is calculated by summing both types of uncertainty (aleatoric and epistemic), obtained from the ensemble method:
\begin{eqnarray}
\label{ann1}
    \Delta\sigma^2_{\text{final}}=\underbrace{\frac{1}{N}\sum_{i=1}^{N}\Delta\sigma^2_{\text{model},i}}_{\text{Aleatoric}}+\underbrace{\frac{1}{N}\sum_{i=1}^{N}(\mu_i-\mu_{\text{final}})^2}_{\text{Epistemic}},
\end{eqnarray}
where the first term of Eq.~(\ref{ann1}) is the mean of the aleatoric uncertainty predicted by each model and the second term of Eq.~(\ref{ann1}) is the epistemic uncertainty, quantitatively measured as the variance (disagreement) of the $N$ models' central value predictions. This final error bar ($\Delta\sigma_{\text{final}} = \sqrt{\Delta\sigma^2_{\text{final}}}$) is what we used in all result plots, as it represents the total uncertainty (data + model) of the ensemble prediction.

\begin{table}[htbp]
\centering
\caption{Detailed specifications of the Deep Ensemble Neural Network architecture and training hyperparameters.}
\label{tab:ann_architecture}
\begin{tabular}{ll}
\toprule
\textbf{Hyperparameter} & \textbf{Configuration/Value} \\
\midrule
\multicolumn{2}{l}{\textit{Network Architecture}} \\
Input Features & 4 ($\log Q^2, \log W, t, \log Y$) \\
Hidden Layers & 3 Dense Layers ($64, 64, 32$ neurons) \\
Output Nodes & 2 ($\mu, \log\sigma^2$) \\
Hidden Activation & Tanh \\
Output Activation & Linear (features $\log$-variance for numerical stability) \\
Weight Initialization & Random (Seed-dependent per ensemble member) \\
\midrule
\multicolumn{2}{l}{\textit{Training Strategy}} \\
Ensemble Size & $N=100$ independent models \\
Loss Function & Heteroscedastic Gaussian NLL \\
Optimizer & Adam \\
Learning Rate & $3 \times 10^{-4}$ \\
Regularization & L2 Norm ($\lambda = 10^{-4}$) applied to kernels \\
Batch Size & 32 \\
Max Epochs & 25,000 \\
Stopping Criteria & Early Stopping (Patience = 1,000 epochs) \\
Validation Split & Fixed Stratified Split (10\% Test, 10\% Val from Train) \\
\bottomrule
\end{tabular}
\end{table}

\subsection{Training and Evaluation Metrics}
\begin{figure}[ht]
    \centering
    \includegraphics[width=0.75\linewidth]{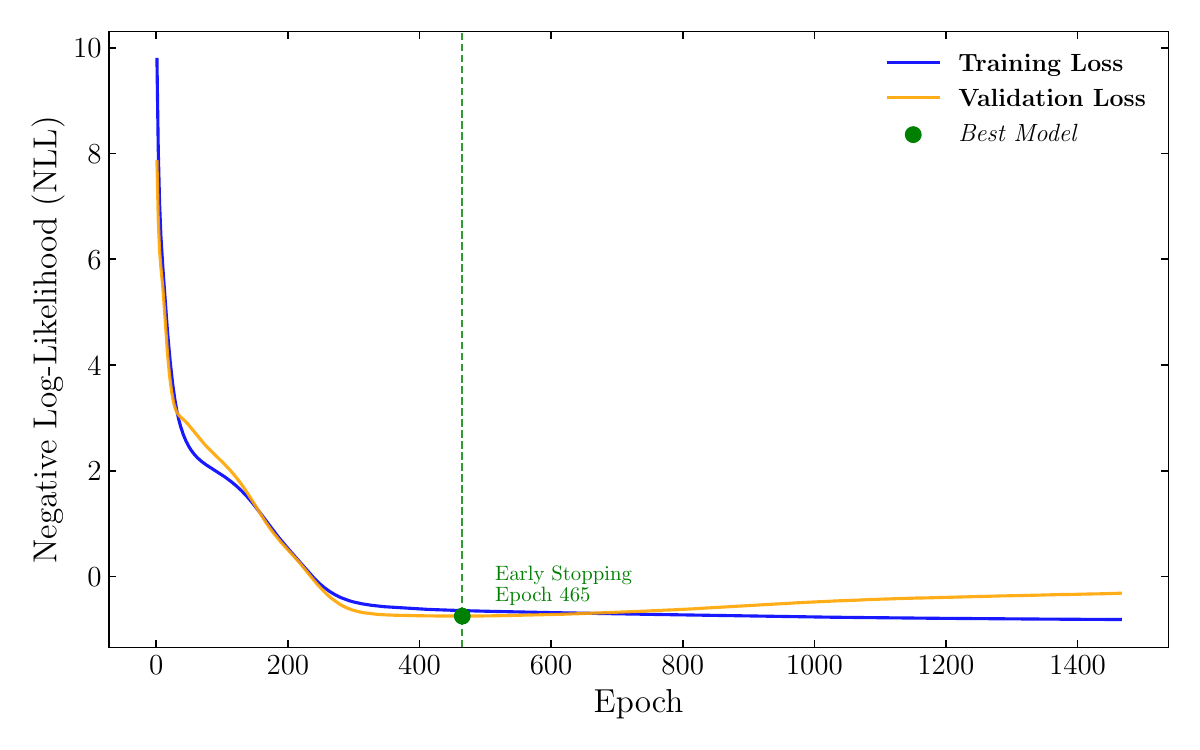}
    \caption{Results for the NLL for the training and validation losses as a function of epoch.} 
    \label{fig:nlllos}
\end{figure}

Each model in the $N=100$ ensemble is trained individually. The optimization is performed using the Adam optimizer~\cite{Kingma:2014vow} with a learning rate of $3 \times 10^{-4}$ and a batch size of 32. Although the maximum number of epochs is set to 25,000, an \texttt{EarlyStopping} callback was implemented to prevent overfitting. This callback monitors the loss on the validation set (\texttt{val\_loss}). If no improvement is observed for 1,000 consecutive epochs (patience), training is halted, and the weights from the epoch with the best \texttt{val\_loss} are restored.
Figure~\ref{fig:nlllos} displays the training and validation loss history for a representative model from the ensemble, demonstrating typical convergence behavior where the validation loss stabilizes alongside the training loss, effectively avoiding overfitting.

\begin{figure}[ht]
    \centering
    \includegraphics[width=0.75\linewidth]{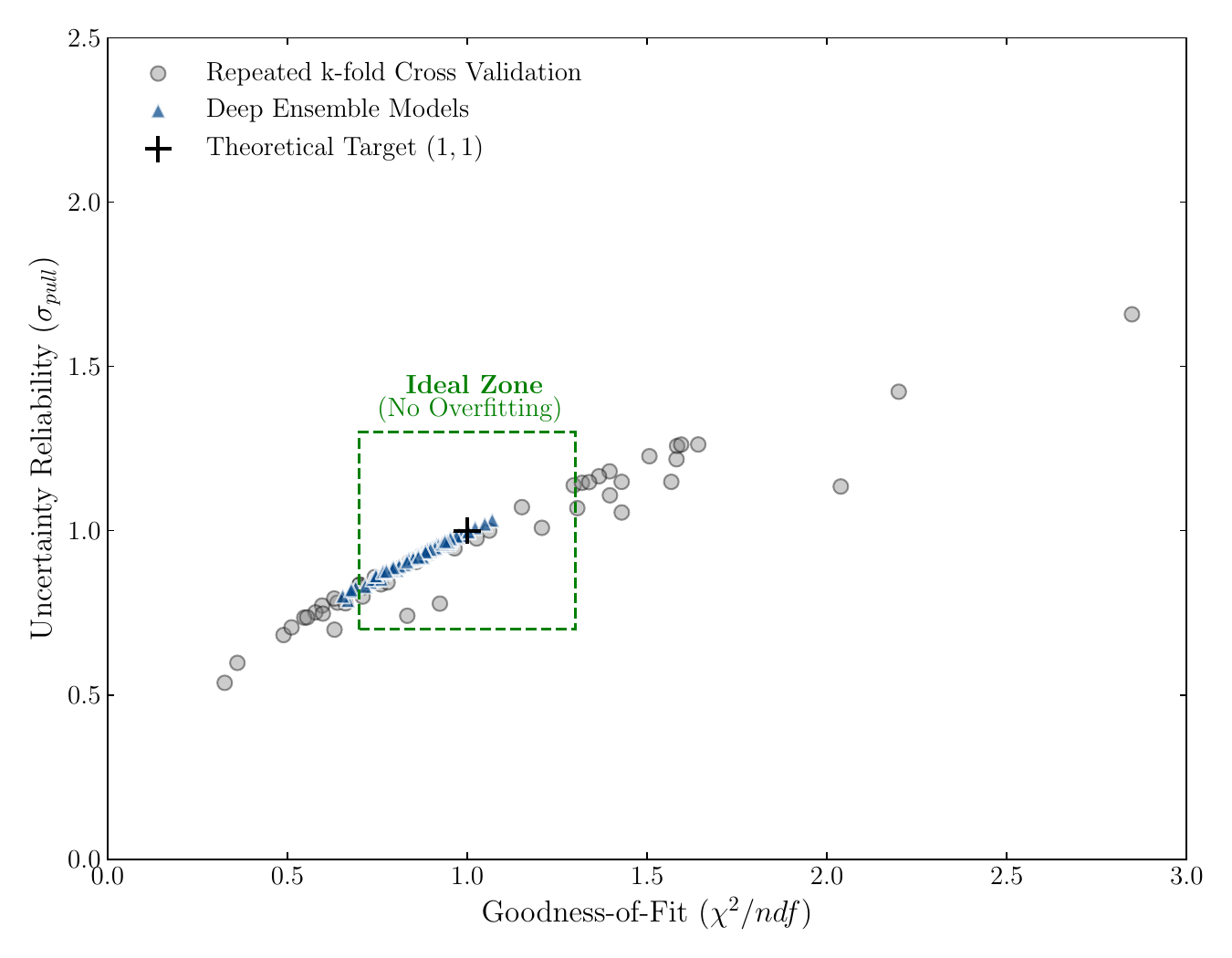}
    \caption{Stability and reliability comparison between Repeated $k$-fold Cross Validation results (gray circles) and the Deep Ensemble models (blue triangles) on the test set.} 
    \label{fig:overfitting_avoidance}
\end{figure}

The performance of the final ensemble is evaluated on the test set (10\% of the data), which is never seen during training. Two primary statistical metrics used for validation are
\begin{enumerate}
    \item \textbf{Chi-squared per Degree of Freedom ($\chi^2 / \text{ndf}$)}. This metric evaluates the goodness-of-fit between the model prediction $\mu_{\text{final}}$ and the data $\bar{y}_{\text{true}}$, taking into account the combined error bar $\Delta\sigma_{\text{final}}$ (which encompasses both $\Delta\sigma_{\text{exp}}$ and $\Delta\sigma_{\text{model}}$). A value of $\chi^2/\text{ndf} \approx 1.0$ indicates that the predicted error bars are consistent with the data dispersion.
    \item \textbf{Pull Distribution}. This metric normalizes the residuals and is defined as $\text{Pull} = (\bar{y}_{\text{true}} - \mu_{\text{final}})/ \Delta\sigma_{\text{final}}$. If the central value predictions and their error bars are accurate, the Pull distribution should follow a standard Gaussian distribution $\mathcal{N}(0,1)$, where $\mu_{\text{pull}}\approx 0$ and $\sigma_{\text{pull}}\approx 1$.
\end{enumerate}
Based on the evaluation of the 100 models on the test set, our combined ensemble achieved robust and stable performance, with average values of $\chi^2/\text{ndf} = 0.86 \pm 0.08$, $\mu_{\text{pull}}= -0.10 \pm 0.04$, and $\sigma_{\text{pull}} = 0.92 \pm 0.05$. These results indicate that the model is not significantly biased (Pull Mean $\approx 0$) and that the predicted error bars (combined aleatoric + epistemic) are slightly conservative but highly accurate (Pull Std $\approx 1$). To visually verify this generalization capability, we present a stability analysis in Fig.~\ref{fig:overfitting_avoidance}.

Figure~\ref{fig:overfitting_avoidance} shows a reliability map for the model's generalization capability. The gray circles illustrate the high instability observed during preliminary tests using repeated $k$-fold cross-validation, where the performance fluctuates significantly, and 
the blue triangles represent the 100 individual models of our final Deep Ensemble, which form a tight cluster. 
This clearly indicates that the cluster falls within the ideal reliability zone (green dashed box), centered at the theoretical 
target of $\chi^2/\text{ndf} \approx 1$ and $\sigma_{\text{pull}} \approx 1$. Such convergence confirms that the ensemble method successfully propagates both experimental and model uncertainties, resulting in predictions that are statistically consistent with the data precision without overfitting.

\section{Numerical result and discussion} \label{sec:IV}
\begin{figure}[ht]
    \centering
    \includegraphics[width=0.75\linewidth]{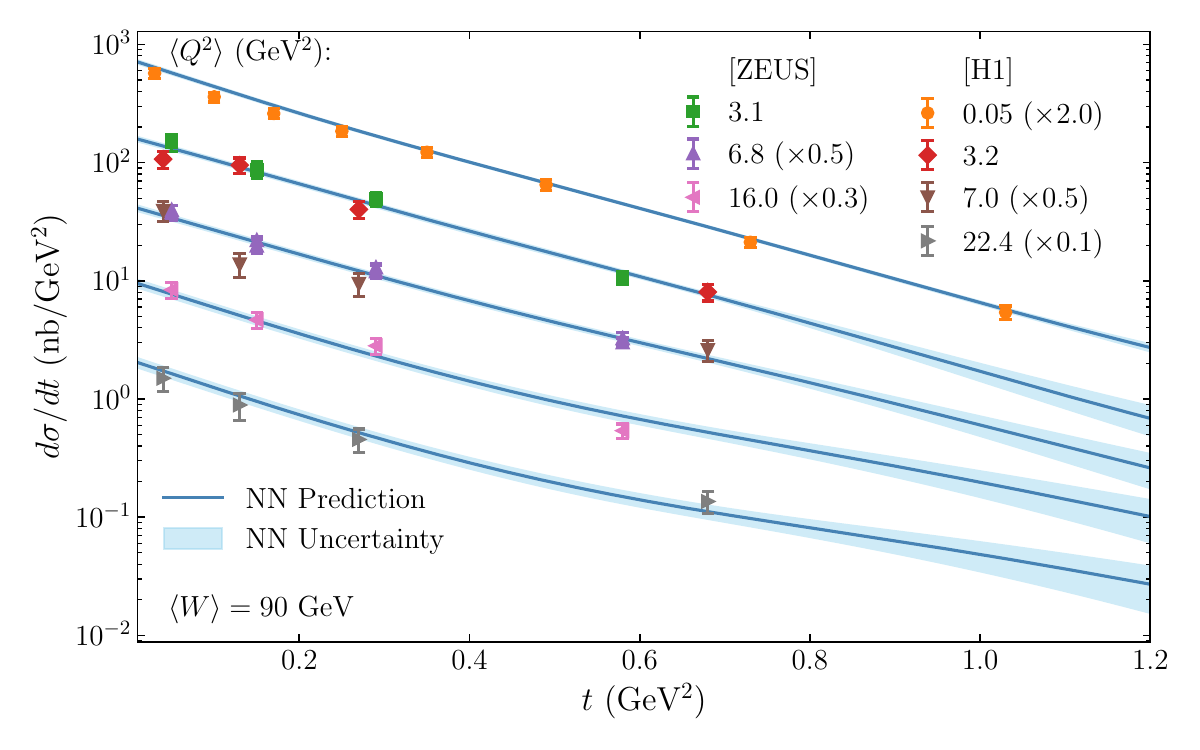}
    \caption{Differential cross section of the exclusive coherent diffractive $J/\psi$ production at fixed $\big<W \big> =$ 90 GeV with different $\big<Q^2\big>$ in comparison with the H1 dan ZEUS data as a function of $t$.} 
    \label{fig1}
\end{figure}

\begin{figure}[htb]
    \centering
    \includegraphics[width=0.75\linewidth, keepaspectratio]{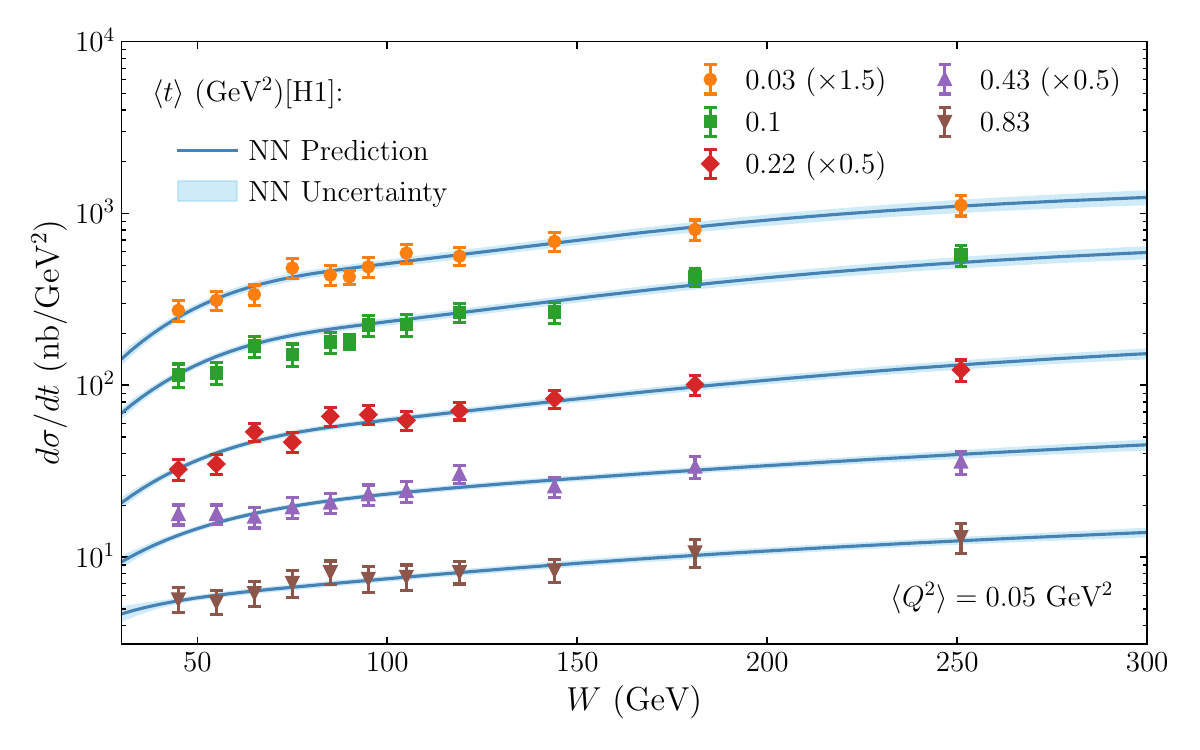}
    \caption{Differential cross section of the exclusive coherent diffractive $J/\Psi$ production at fixed $\big<Q^2\big> =$ 0.05 GeV$^2$ for different $t$ as a function of $W$.}
    \label{fig2}
\end{figure}

\begin{figure}[htb]
    \centering
    \includegraphics[width=0.75\linewidth, keepaspectratio]{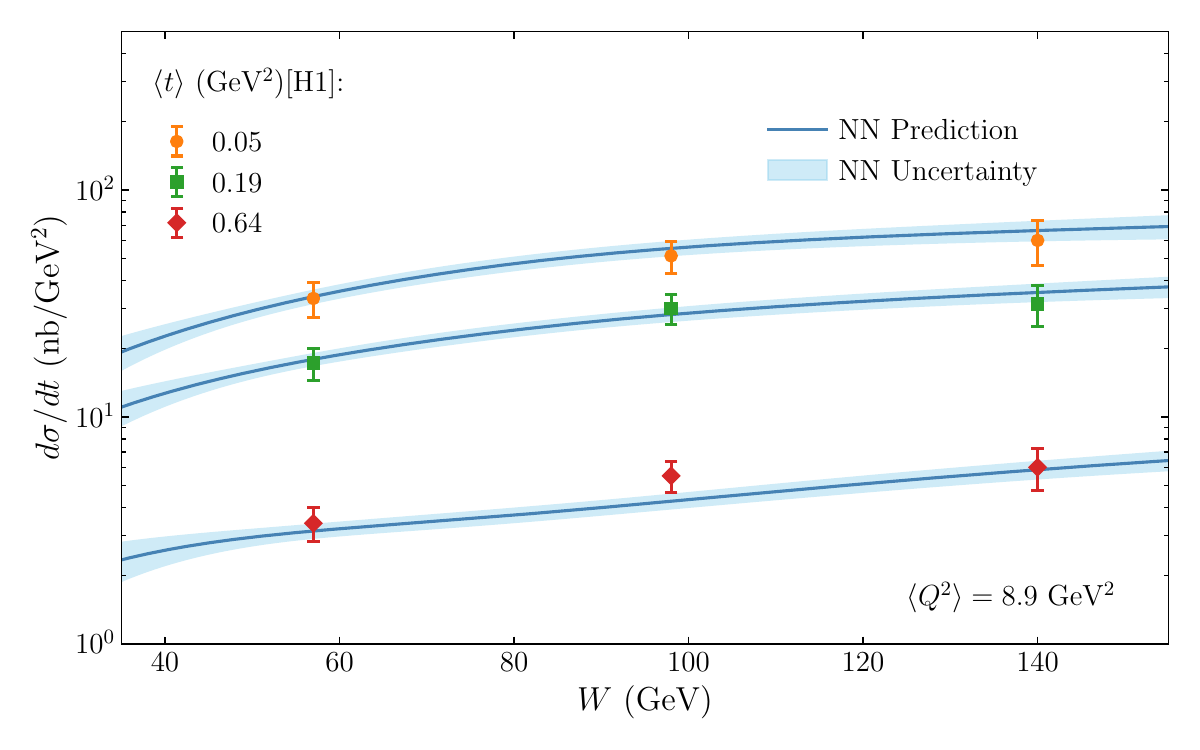}
    \caption{Center of mass energy ($W$) dependence of the differential cross-section of the exclusive $J/\psi$ electroproduction at fixed $\big<Q^2\big> =$ 8.9 GeV$^2$ for the values of $t=$ 0.05, 0.19, and 0.64 GeV$^2$.}
    \label{fig3}
\end{figure}

\begin{figure}[ht]
    \centering
    \includegraphics[width=0.75\linewidth, keepaspectratio]{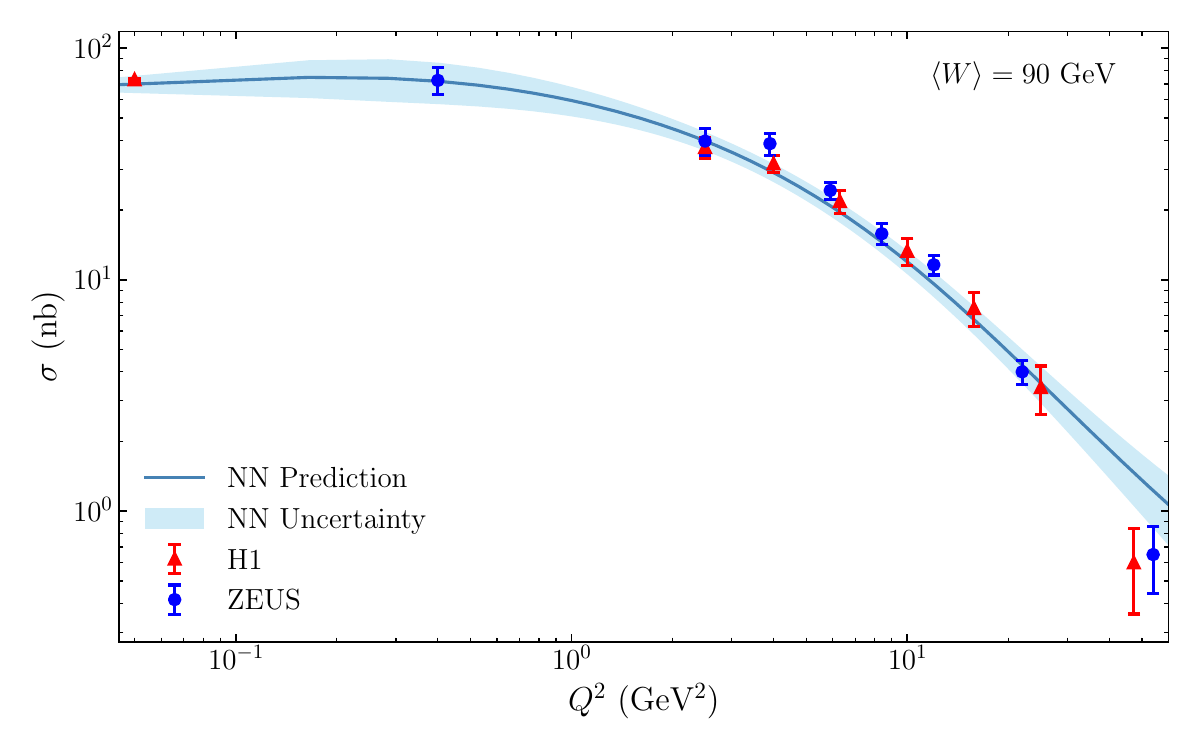}
    \caption{Total cross-section of $\gamma^{(*)} p \rightarrow J/\psi \, p$ at fixed $W=$ 90 GeV as a function of $Q^2$, in comparison with H1 (triangle point) and ZEUS (circle point) data. The experimental data of the H1 and ZEUS collaborations are taken from Refs.~\cite{H1:2005dtp,ZEUS:2004yeh}.}
    \label{fig4}
\end{figure}

\begin{figure}[htb]
    \centering
    \includegraphics[width=0.75\linewidth, keepaspectratio]{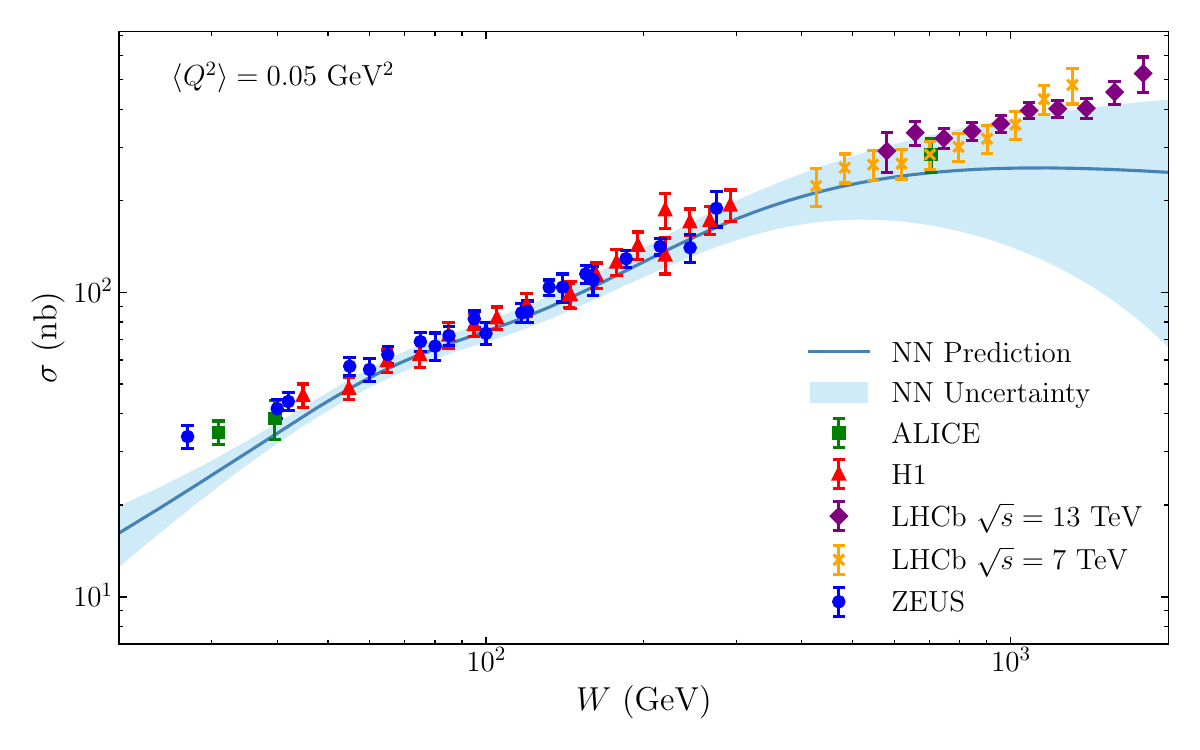}
    \caption{Total cross-section of the $\gamma p \rightarrow J/\psi p$ at fixed $\big<Q^2 \big> =$ 0.05 GeV$^2$ as a function of $W$, in comparison with ALICE data (square point), H1 (triangle point), LHCb $\sqrt{s} =$ 13 TeV (diamond point), LHCb $\sqrt{s}=$ 7 TeV (cross point), and ZEUS (circle point) data. Data are taken from Ref.~\cite{LHCb:2018rcm}}.
    \label{fig5}
\end{figure}

\begin{figure}[ht]
    \centering
    \includegraphics[width=0.55\linewidth, keepaspectratio]{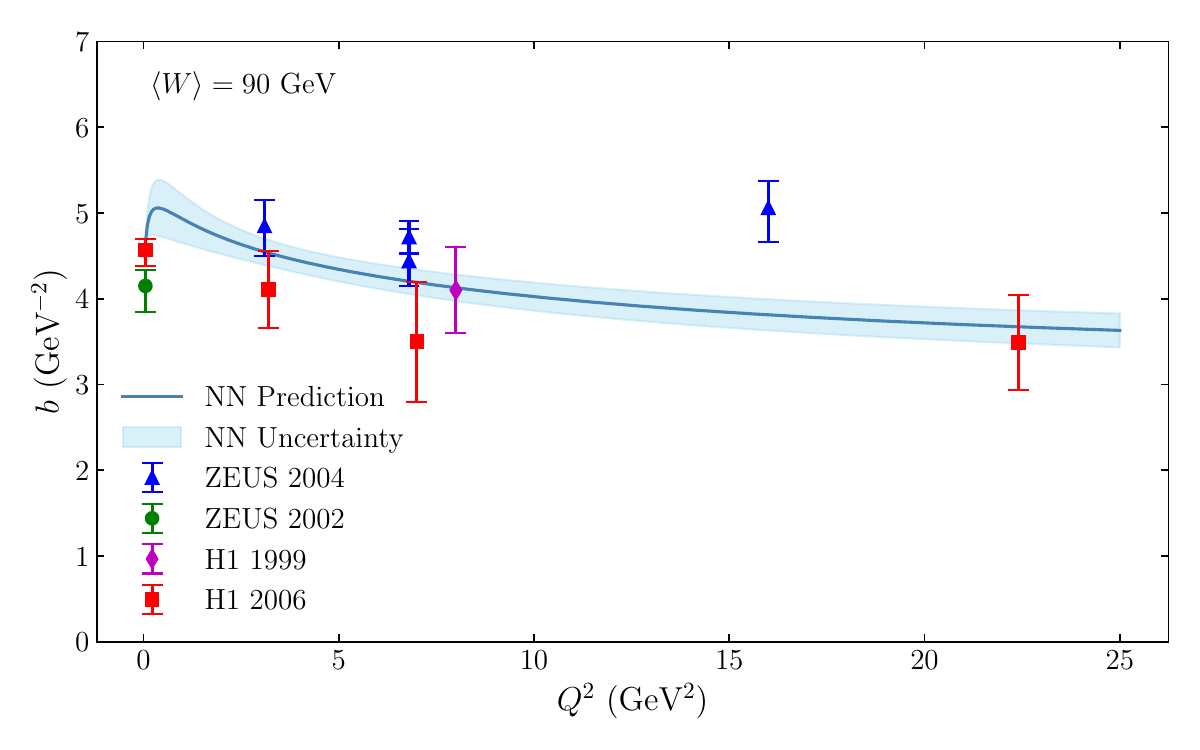} 
    \includegraphics[width=0.475\linewidth, keepaspectratio]{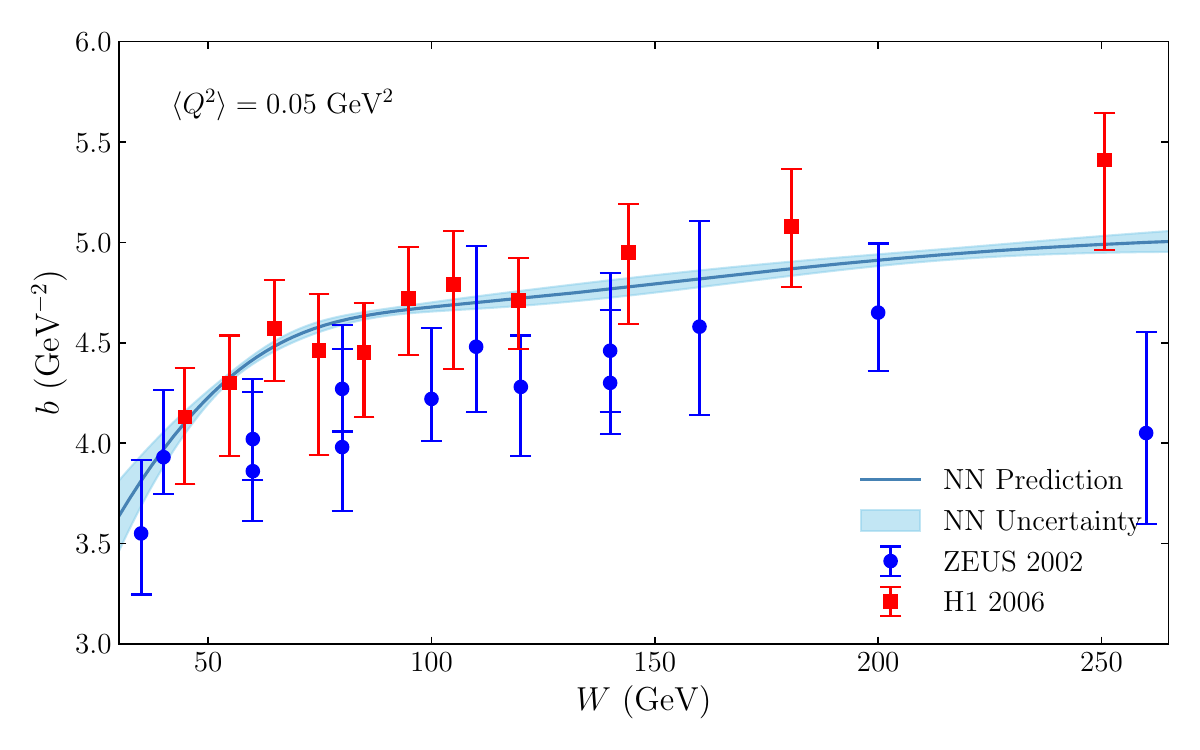}
    \includegraphics[width=0.475\linewidth, keepaspectratio]{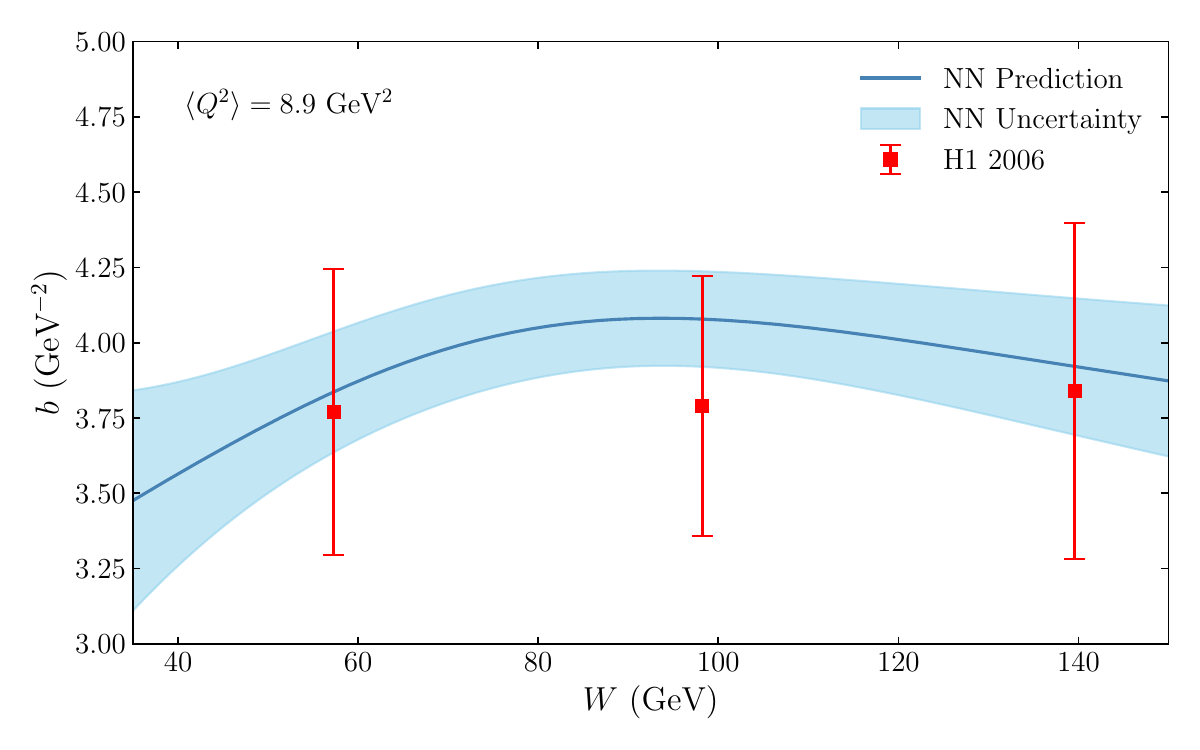}
    \caption{Results for the $t$-slope parameter $b$ extracted from the predictive ANN model for $J/\psi$ photoproduction at $W =$ 90 GeV as a function of $Q^2$ (upper center panel), $b$ as a function of $W$ for $\big<Q^2\big> =$ 0.05 GeV$^2$ (lower left panel), and $b$ as a function of $W$ for $\big<Q^2\big> =$ 8.9 GeV$^2$ (lower right panel) in comparison with the ZEUS 2004 (triangle point), ZEUS 2002 (circle point), H1 1999 (diamond point), and H1 2006 (square point) data.}
    \label{fig6}
\end{figure}

Our data-driven ANN model prediction results of the $t$- and $W$-dependencies of the differential and total cross-section for the exclusive coherent diffractive $J/\psi$ production for the HERA, LHC, and EIC data are presented. Results of our ANN model prediction for the differential and total cross-section of the exclusive coherent diffractive $J/\psi$ production and results for the extraction of the $t$-slope parameter $b$ are depicted in Figs.~\ref{fig1}-\ref{fig6}.

Fig.~\ref{fig1} shows the ANN model prediction for the differential cross-section $d\sigma/dt$ at fixed values of $ W=$90 GeV for different values of $Q^2$ as a function of $t$. This indicates that our ANN models for the $t$-dependence of the differential cross-sections of diffractive $J/\psi$ production decrease as $t$ and $Q^2$ increase, which is consistent with the data for the corresponding $Q^2$ values, except the $d\sigma/dt$ for the $\big<Q^2\big> =$ 3.2 and 7.0 GeV$^2$ at $t<0.3$ GeV$^2$, our ANN models rather underestimate the data. However, overall, our ANN model predictions are well described by the data for all values of $Q^2$. It is worth noting that we do not make any theoretical assumptions or constraints in determining our ANN models, as already explained earlier in Sec.~\ref{sec:III}.  

In Fig.~\ref{fig2}, we show our results for the ANN model predictions of $W$-dependence of the $J/\psi$ differential cross section $d\sigma/dt$ at fixed values of $\big<Q^2\big> =$ 0.05 GeV$^2$ for different values of $t$. We find that the ANN models well describe the data for almost all values of $t$. However, a few data points for $ t=$0.1, and 0.43 GeV$^2$ at around $W \lesssim$ 100 GeV do not fit the data. We expect that this is due to the uncertainties at those data points being large, relative to other data points. Also, we find that the differential cross-section of our ANN model slowly increases as the $W$ increases, and it increases with decreasing $t$.

In addition to our ANN model prediction results in Fig.~\ref{fig2}, we also obtain the ANN models for the $W$-dependence differential cross section of the exclusive coherent $J/\psi$ production at $\big<Q^2 \big>=$ 8.9 GeV$^2$ (higher than the $\big<Q^2\big>$ used in Fig.~\ref{fig2}) for values of $t=$ 0.05, 0.19, and 0.64 GeV$^2$, as shown in Fig.~\ref{fig3}. It shows that the ANN model prediction is in good agreement with the data for all values of $t$. The trend of the ANN model for differential cross-section slowly increases as $W$ increases, which is the same as indicated in Fig.~\ref{fig2}. Analogously, a similar trend and behavior is also found for the ANN model for the $t$-dependence of the differential cross-section.

Besides the differential cross-section for the exclusive coherent $J/\psi$ production, we also evaluate our ANN models for the total cross-section. Our ANN model results for the total cross-section at fixed $W=$ 90 GeV as a function of $Q^2$ are illustrated in Fig.~\ref{fig4}. It indicates that our ANN model is in good agreement with the H1 and ZEUS data, except for the H1 data point at large $Q^2$, where our ANN model overestimates that data point of H1 data. Note that our ANN models are simultaneously determined using the combination of H1 and ZEUS data.

Results for our ANN models for the total cross-section of $\gamma \, p \rightarrow J/\psi p$ at fixed values of $\big<Q^2 \big> =$ 0.05 GeV$^2$ as a function of $W$ are depicted in Fig.~\ref{fig5}. It shows that our ANN models fit the data well at intermediate values of $W$. However, at lower $W$, our ANN models underestimate data points of ALICE (square point) and ZEUS (circle point) data. Similar behavior is also found at higher $W$, where our ANN models underestimate the data point of LHCb $\sqrt{s} =$ 13 TeV. Moreover, our ANN models exhibit a large uncertainty at around $W \gtrsim 10^3$ GeV. The large uncertainty at large $W$ is expected because the LHCb data at $W > 250~\mathrm{GeV}$ lie outside the kinematic range of the training dataset. As described in Sec.~\ref{data preprocessing}, the ANN is trained exclusively on HERA data covering the range
$20 < W < 250~\mathrm{GeV}$. Note that the total cross section shown in Fig.~\ref{fig5} is obtained by integrating the predicted differential cross section $d\sigma^{\gamma p \rightarrow J/\psi\, p}/dt$ over the range $0 < t < 1.2~\mathrm{GeV}^2$ and evaluating it in the interval
$20 < W < 1500~\mathrm{GeV}$, which extends beyond the region directly constrained by the training data.

Finally, our results for the $b$ slope parameter extracted from the predictive ANN model for $J/\psi$ photoproduction at fixed values of $W=$ 90 GeV as a function of $Q^2$, together with experimental data, are shown in the upper center of Fig.~\ref{fig6}. It shows that our $b$ slope parameter extracted from the ANN models is in good agreement with the ZEUS 2004 (triangle point), and H1 1999 (diamond point) data in the range 2 GeV$^2$ $\lesssim Q^2 \lesssim$ 10 GeV$^2$, and H1 2006 (square point) in the range 2 GeV$^2$ $\lesssim Q^2 \lesssim$ 23 GeV$^2$. Surprisingly, we find that the $b$ slope parameter overestimates the ZEUS 2002 data (circle point) at lower $Q^2$ and underestimates the ZEUS 2004 data (triangle point) at intermediate $Q^2$. However, overall, the slope parameter obtained from our ANN models is slightly consistent with that observed in the $J/\psi$ electroproduction~\cite{ZEUS:2004yeh}, giving the mean value of $b =$ 4.72 $\pm$ 0.15 (stat.) $\pm$ 0.12 (syst.) GeV$^{-2}$ (in the range $2<Q^2<100\, \textrm{GeV}^2$).

Results for the $b$ slope parameter for $\big< Q^2\big> =$ 0.05 and 8.9 GeV$^2$ as a function of $W$ are depicted in the lower panel of Fig.~\ref{fig6}, respectively. It shows that the $b$ slope parameter for $\big<Q^2\big> =$ 0.05 GeV$^2$ is consistent with H1 2002 data, rather than that with ZEUS 2002 data, as in the lower left panel of Fig.~\ref{fig6}. For $b$ at $\big<Q^2\big> = $ 0.05 GeV$^2$, it indicates that $b$ slowly increases as $W$ increases. In the lower right panel of Fig.~\ref{fig6}, it is shown that the $b$ parameter is consistent with the H1 2006 data. It is worth noting that the behavior of the $b$ slope for $\big<Q^2\big> =$ 0.05 GeV$^2$ is similar to that for $\big<Q^2\big> =$ 8.9 GeV$^2$ in the appropriate range of $W$.

\section{Summary and conclusion} \label{sec:V}
In summary, we have developed a data-driven ANN framework to model the exclusive coherent diffractive $J/\psi$ production cross section using the combined HERA dataset. Unlike traditional theoretical approaches based on the dipole picture, which rely on multiple model-dependent ingredients and are often constrained to limited kinematic regions, the ANN method offers a flexible, assumption-minimized alternative that can capture nonlinear correlations in the data.

Using the processed H1 and ZEUS differential cross-section measurements, we trained an ensemble of ANN models incorporating both aleatoric and epistemic uncertainties. The resulting ensemble achieves stable performance, with $\chi^2/\mathrm{ndf} \approx 1$ and pull distributions consistent with a standard Gaussian, indicating that both the central predictions and uncertainty estimates are reliable.

The ANN predictions for the $t$- and $W$-dependence of $d\sigma/dt$ show good agreement with the HERA data across a broad range of $Q^2$ and $t$. We also extended the model to predict the total photoproduction cross section by combining HERA and LHC measurements, observing reasonable consistency over most of the accessible $W$ range despite limited high-energy data. The extracted exponential slope parameter $b$ agrees well with the experimental data.

Overall, we concluded that the ANN framework presented here demonstrates that data-driven approaches could complement traditional QCD-based models by reducing sensitivity to specific theoretical assumptions and providing robust interpolation across multi-dimensional kinematic regions. With the present results, we plan to extend this method to a larger dataset in the future. In this context, another interesting direction for future work is the development of hybrid or physics-informed ANN frameworks, in which the ANN is trained alongside theoretical predictions in kinematic regions where they are reliable. Such approaches could improve extrapolation beyond the available data, as demonstrated in Ref.~\cite{Kou:2026iau}, where Physics-Informed Neural Networks embedding QCD evolution dynamics could achieve universal predictions for exclusive processes. Also, recent global Bayesian analyses of $J/\psi$ photoproduction have highlighted the difficulty of achieving a simultaneous description of proton and nuclear data within fixed theoretical frameworks \cite{Mantysaari:2025ltq}, suggesting that theoretical consistency constraints should also be considered in future developments. Hence, we aim to include additional data from other vector mesons in our future analysis, which may provide further insight into small-$x$ saturation dynamics and the proton transverse profile. Additionally, the method used in recent studies will be a valuable tool to employ for analyzing the high-precision vector meson data to be measured at future and modern facilities~\cite{Accardi:2012qut,Accardi:2023chb}.

\section*{Acknowledgments}
We thank the National Research and Innovation Agency (BRIN), Republic of Indonesia, for providing the Mahameru High Performance Computing Facility, where most of the computation in this paper was performed. P.T.P.H. was partially supported by the RCNP Collaboration Research Network program under project number COREnet 057. The work of C.S. was supported by Hibah Penelitian FMIPA UGM Tahun Anggaran 2025.

\section*{ORCID}
\noindent Taufiq~I.~Baihaqi\orcid{0009-0008-3148-6994}   \url{https://orcid.org/0009-0008-3148-6994} 

\noindent Zulkaida Akbar\orcid{0000-0002-5373-6121}   \url{https://orcid.org/0000-0002-5373-6121} 

\noindent Chalis Setyadi\orcid{0000-0002-5853-4238} \url{https://orcid.org/0000-0002-5853-4238}

\noindent Parada~T.~P.~Hutauruk\orcid{0000-0002-4225-7109}  \url{https://orcid.org/0000-0002-4225-7109}

\noindent Apriadi Salim Adam\orcid{0000-0001-6587-5156} \url{https://orcid.org/0000-0001-6587-5156}



\begin{thebibliography}{0}
\bibitem{Lepage:1980fj}
G.~P.~Lepage and S.~J.~Brodsky,
\href{https://doi.org/10.1103/PhysRevD.22.2157}{Phys. Rev. D \textbf{22}, 2157 (1980)}.

\bibitem{Accardi:2012qut}
A.~Accardi, J.~L.~Albacete, M.~Anselmino, N.~Armesto, E.~C.~Aschenauer, A.~Bacchetta, D.~Boer, W.~K.~Brooks, T.~Burton and N.~B.~Chang, \textit{et al.}
\href{https://doi.org/10.1140/epja/i2016-16268-9}{Eur. Phys. J. A \textbf{52}, no.9, 268 (2016)}.

\bibitem{Li:2022kwn}
X.~Li, J.~K.~Adkins, Y.~Akiba, A.~Albataineh, M.~Amaryan, I.~C.~Arsene, C.~Ayerbe Gayoso, J.~Bae, X.~Bai and M.~D.~Baker, \textit{et al.}
\href{https://doi.org/10.1016/j.nima.2022.167956}{Nucl. Instrum. Meth. A \textbf{1048}, 167956 (2023)}.

\bibitem{Anderle:2021wcy}
D.~P.~Anderle, V.~Bertone, X.~Cao, L.~Chang, N.~Chang, G.~Chen, X.~Chen, Z.~Chen, Z.~Cui and L.~Dai, \textit{et al.}
\href{https://doi.org/10.1007/s11467-021-1062-0}{Front. Phys. (Beijing) \textbf{16}, no.6, 64701 (2021)}.

\bibitem{LHeCStudyGroup:2012zhm}
J.~L.~Abelleira Fernandez \textit{et al.} [LHeC Study Group],
\href{https://doi.org/10.1088/0954-3899/39/7/075001}{J. Phys. G \textbf{39}, 075001 (2012)}.

\bibitem{Ryskin:1992ui}
M.~G.~Ryskin,
\href{https://doi.org/10.1007/BF01555742}{Z. Phys. C \textbf{57}, 89-92 (1993)}.

\bibitem{Brodsky:1994kf}
S.~J.~Brodsky, L.~Frankfurt, J.~F.~Gunion, A.~H.~Mueller and M.~Strikman,
\href{https://doi.org/10.1103/PhysRevD.50.3134}
{Phys. Rev. D \textbf{50}, 3134-3144 (1994)}

\bibitem{Frankfurt:1995jw}
L.~Frankfurt, W.~Koepf and M.~Strikman,
\href{https://doi.org/10.1103/PhysRevD.54.3194}
{Phys. Rev. D \textbf{54}, 3194-3215 (1996)}

\bibitem{Collins:1996fb}
J.~C.~Collins, L.~Frankfurt and M.~Strikman,
\href{https://doi.org/10.1103/PhysRevD.56.2982}
{Phys. Rev. D \textbf{56}, 2982-3006 (1997)}

\bibitem{Boer:2024ylx}
D.~Boer, C.~A.~Flett, C.~Flore, D.~Kiko{\l}a, J.~P.~Lansberg, M.~Nefedov, C.~Van Hulse, S.~Bhattacharya, J.~Bor and M.~Butenschoen, \textit{et al.}
\href{https://doi.org/10.1016/j.ppnp.2025.104162}{Prog. Part. Nucl. Phys. \textbf{142}, 104162 (2025)}.

\bibitem{H1:2000kis}
C.~Adloff \textit{et al.} [H1],
\href{https://doi.org/10.1016/S0370-2693(00)00530-X}{Phys. Lett. B \textbf{483} (2000), 23-35}.

\bibitem{H1:2005dtp}
A.~Aktas \textit{et al.} [H1],
\href{https://doi.org/10.1140/epjc/s2006-02519-5}{Eur. Phys. J. C \textbf{46} (2006), 585-603}.

\bibitem{H1:2013okq}
C.~Alexa \textit{et al.} [H1],
\href{https://doi.org/10.1140/epjc/s10052-013-2466-y}{Eur. Phys. J. C \textbf{73} (2013) no.6, 2466}.

\bibitem{ZEUS:2002wfj}
S.~Chekanov \textit{et al.} [ZEUS],
\href{https://doi.org/10.1007/s10052-002-0953-7}{Eur. Phys. J. C \textbf{24} (2002), 345-360}.


\bibitem{Golec-Biernat:1999qor}
K.~J.~Golec-Biernat and M.~Wusthoff,
\href{https://doi.org/10.1103/PhysRevD.60.114023}{Phys. Rev. D \textbf{60} (1999), 114023}

\bibitem{Golec-Biernat:1998zce}
K.~J.~Golec-Biernat and M.~Wusthoff,
\href{https://doi.org/10.1103/PhysRevD.59.014017}{Phys. Rev. D \textbf{59} (1998), 014017}


\bibitem{Gelis:2010nm}
F.~Gelis, E.~Iancu, J.~Jalilian-Marian and R.~Venugopalan,
\href{https://doi.org/10.1146/annurev.nucl.010909.083629}{Ann. Rev. Nucl. Part. Sci. \textbf{60}, 463-489 (2010)}.

\bibitem{Kowalski:2003hm}
H.~Kowalski and D.~Teaney,
\href{https://doi.org/10.1103/PhysRevD.68.114005}{Phys. Rev. D \textbf{68}, 114005 (2003)}.

\bibitem{Weigert:2005us}
H.~Weigert,
\href{https://doi.org/10.1016/j.ppnp.2005.01.029}{Prog. Part. Nucl. Phys. \textbf{55}, 461-565 (2005)}.

\bibitem{Amoroso:2022eow}
S.~Amoroso, A.~Apyan, N.~Armesto, R.~D.~Ball, V.~Bertone, C.~Bissolotti, J.~Bluemlein, R.~Boughezal, G.~Bozzi and D.~Britzger, \textit{et al.}
\href{https://doi.org/10.5506/APhysPolB.53.12-A1}{Acta Phys. Polon. B \textbf{53}, no.12, 12-A1 (2022)}.

\bibitem{Schlichting:2014ipa}
S.~Schlichting and B.~Schenke,
\href{doi:10.1016/j.physletb.2014.10.068}{Phys. Lett. B \textbf{739}, 313-319 (2014)}.

\bibitem{Mantysaari:2016ykx}
H.~M{\"a}ntysaari and B.~Schenke,
\href{https://doi.org/10.1103/PhysRevLett.117.052301}{Phys. Rev. Lett. \textbf{117}, no.5, 052301 (2016)}.

\bibitem{Mantysaari:2016jaz}
H.~M{\"a}ntysaari and B.~Schenke,
\href{https://doi.org/10.1103/PhysRevD.94.034042}{Phys. Rev. D \textbf{94}, no.3, 034042 (2016)}.



\bibitem{H1:1996prv}
S.~Aid \textit{et al.} [H1],
\href{https://doi.org/10.1016/0550-3213(96)00045-4}{Nucl. Phys. B \textbf{463}, 3-32 (1996)}.

\bibitem{ZEUS:1995bfs}
M.~Derrick \textit{et al.} [ZEUS],
\href{https://doi.org/10.1007/s002880050004}{Z. Phys. C \textbf{69}, 39-54 (1995)}.

\bibitem{Donnachie:1994zb}
A.~Donnachie and P.~V.~Landshoff,
\href{https://doi.org/10.1016/0370-2693(95)00115-2}{Phys. Lett. B \textbf{348}, 213-218 (1995)}.

\bibitem{ZEUS:2004yeh}
S.~Chekanov \textit{et al.} [ZEUS],
\href{https://doi.org/10.1016/j.nuclphysb.2004.06.034}{Nucl. Phys. B \textbf{695}, 3-37 (2004)}.

\bibitem{Bursche:2018eni}
A.~Bursche [LHCb],
\href{https://doi.org/10.1016/j.nuclphysa.2018.10.069}{Nucl. Phys. A \textbf{982}, 247-250 (2019)}.

\bibitem{LHCb:2018rcm}
R.~Aaij \textit{et al.} [LHCb],
\href{https://doi.org/10.1007/JHEP10(2018)167}{HEP \textbf{10}, 167 (2018)}.

\bibitem{Iancu:2017fzn}
E.~Iancu and A.~H.~Rezaeian,
\href{https://doi.org/10.1103/PhysRevD.95.094003}{Phys. Rev. D \textbf{95}, no.9, 094003 (2017)}.

\bibitem{Mueller:1994jq}
A.~H.~Mueller and B.~Patel,
\href{https://doi.org/10.1016/0550-3213(94)90284-4}{Nucl. Phys. B \textbf{425}, 471-488 (1994)}

\bibitem{Kowalski:2006hc}
H.~Kowalski, L.~Motyka and G.~Watt,
\href{https://doi.org/10.1103/PhysRevD.74.074016}{Phys. Rev. D \textbf{74}, 074016 (2006)}.

\bibitem{Albertsson:2018maf}
K.~Albertsson, P.~Altoe, D.~Anderson, J.~Anderson, M.~Andrews, J.~P.~Araque Espinosa, A.~Aurisano, L.~Basara, A.~Bevan and W.~Bhimji, \textit{et al.}
\href{https://doi.org/10.1088/1742-6596/1085/2/022008}{J. Phys. Conf. Ser. \textbf{1085}, no.2, 022008 (2018)}.

\bibitem{Boer:2023mip}
D.~Boer and C.~Setyadi,
\href{https://doi.org/10.1140/epjc/s10052-023-12040-6}{Eur. Phys. J. C \textbf{83}, no.10, 890 (2023)}.

\bibitem{Dosch:1996ss}
H.~G.~Dosch, T.~Gousset, G.~Kulzinger and H.~J.~Pirner,
\href{https://doi.org/10.1103/PhysRevD.55.2602}{Phys. Rev. D \textbf{55}, 2602-2615 (1997)}

\bibitem{Nemchik:1996cw}
J.~Nemchik, N.~N.~Nikolaev, E.~Predazzi and B.~G.~Zakharov,
\href{https://doi.org/10.1007/s002880050448}{Z. Phys. C \textbf{75}, 71-87 (1997)}.

\bibitem{Kopeliovich:1991pu}
B.~Z.~Kopeliovich and B.~G.~Zakharov,
\href{https://doi.org/10.1103/PhysRevD.44.3466}{Phys. Rev. D \textbf{44}, 3466-3472 (1991)}.

\bibitem{McLerran:1993ni}
L.~D.~McLerran and R.~Venugopalan,
\href{https://doi.org/10.1103/PhysRevD.49.2233}{Phys. Rev. D \textbf{49}, 2233-2241 (1994)}.

\bibitem{McLerran:1993ka}
L.~D.~McLerran and R.~Venugopalan,
\href{https://doi.org/10.1103/PhysRevD.49.3352}{Phys. Rev. D \textbf{49}, 3352-3355 (1994)}.

\bibitem{Iancu:2003ge}
E.~Iancu, K.~Itakura and S.~Munier,
\href{https://doi.org/10.1016/j.physletb.2004.02.040}{Phys. Lett. B \textbf{590}, 199-208 (2004)}.

\bibitem{Forshaw:2004vv}
J.~R.~Forshaw and G.~Shaw,
\href{https://doi.org/10.1088/1126-6708/2004/12/052}{JHEP \textbf{12}, 052 (2004)}.

\bibitem{Forshaw:2003ki}
J.~R.~Forshaw, R.~Sandapen and G.~Shaw,
\href{https://doi.org/10.1103/PhysRevD.69.094013}{Phys. Rev. D \textbf{69}, 094013 (2004)}.

\bibitem{Martin:1999wb}
A.~D.~Martin, M.~G.~Ryskin and T.~Teubner,
\href{https://doi.org/10.1103/PhysRevD.62.014022}{Phys. Rev. D \textbf{62}, 014022 (2000)}.

\bibitem{Shuvaev:1999ce}
A.~G.~Shuvaev, K.~J.~Golec-Biernat, A.~D.~Martin and M.~G.~Ryskin,
\href{https://doi.org/10.1103/PhysRevD.60.014015}{Phys. Rev. D \textbf{60}, 014015 (1999)}.

\bibitem{Lappi:2012vw}
T.~Lappi and H.~M{\"a}ntysaari,
\href{https://doi.org/10.1140/epjc/s10052-013-2307-z}{Eur. Phys. J. C \textbf{73}, no.2, 2307 (2013)}.


\bibitem{Kendall:2017tnb}
A.~Kendall and Y.~Gal,
\href{https://doi.org/10.48550/arXiv.1703.04977}{[arXiv:1703.04977 [cs.CV]]}.



\bibitem{Chollet:2015}
F.~Chollet {\it et al.},
\url{https://keras.io} (2015).

\bibitem{Krogh:1992}
A.~Krogh and J.~A.~Hertz,
\href{https://proceedings.neurips.cc/paper/1991/hash/8eefcfdf5990e441f0fb6f3fad709e21-Abstract.html}{Adv. Neural Inf. Process. Syst. \textbf{4}, 950-957 (1992)}.


\bibitem{Nix1994EstimatingTM}
D.~A.~Nix and A.~S.~Weigend,
\href{https://doi.org/10.1109/ICNN.1994.374138}{Proc. IEEE Int. Conf. Neural Networks \textbf{1}, 55-60 (1994)}.

\bibitem{Yuan2004DoublyPL}
M.~Yuan and G.~Wahba,
\href{https://doi.org/10.1016/j.spl.2004.03.009}{Statist. Probab. Lett. \textbf{69}, 11-20 (2004)}.

\bibitem{Le2005Heteroscedastic}
Q.~V.~Le, A.~J.~Smola and S.~Canu,
\href{https://dl.acm.org/doi/10.1145/1102351.1102413}{Proc. Int. Conf. Mach. Learn. \textbf{22}, 489-496 (2005)}.

\bibitem{Lakshminarayanan:2016simple}
B.~Lakshminarayanan, \textit{et al.,}
\href{https://papers.nips.cc/paper/7219-simple-and-scalable-predictive-uncertainty-estimation-using-deep-ensembles}{Adv. Neural Inf. Process. Syst. \textbf{30}, 6402-6413 (2017)}.

\bibitem{Kingma:2014vow}
D.~P.~Kingma and J.~Ba,
\href{https://doi.org/10.48550/arXiv.1412.6980}{[arXiv:1412.6980 [cs.LG]]}.

\bibitem{Kou:2026iau}
W.~Kou and X.~Chen,
\href{https://doi.org/10.48550/arXiv.2601.16391}{[arXiv:2601.16391 [hep-ph]]}.

\bibitem{Mantysaari:2025ltq}
H.~M{\"a}ntysaari, H.~Roch, F.~Salazar, B.~Schenke, C.~Shen and W.~Zhao,
\href{https://doi.org/10.1103/pcmz-dyz1}{Phys. Rev. D \textbf{113} (2026) no.1, 014038}.

\bibitem{Accardi:2023chb}
A.~Accardi, P.~Achenbach, D.~Adhikari, A.~Afanasev, C.~S.~Akondi, N.~Akopov, M.~Albaladejo, H.~Albataineh, M.~Albrecht and B.~Almeida-Zamora, \textit{et al.}
\href{https://doi.org/10.1140/epja/s10050-024-01282-x}{Eur. Phys. J. A \textbf{60}, no.9, 173 (2024)}.

\end{thebibliography}
\end{document}